\documentclass{new_tlp}

\usepackage{verbatim}
\usepackage{xspace}
\usepackage{subfigure}
\usepackage{graphicx}
\usepackage{multirow}
\usepackage[latin1]{inputenc}
\usepackage[english]{babel}
\usepackage{amsmath, amsfonts, amssymb}

\newcommand{\Reals}{\ensuremath{\mathbb{R}}}

\newcommand{\Nat}{\ensuremath{\mathbb{N}}}

\newcommand{\Rat}{\ensuremath{\mathbb{Q}}}

\newcommand{\Prob}[1]{\ensuremath{Pr(#1)}}

\newcommand{\Esp}[1]{\ensuremath{\mathbb{E}[#1]}}

\newcommand{\Distrib}[1]{\ensuremath{f_{#1}}}

\newcommand{\cumulative}[1]{\ensuremath{\mathcal{F}_{#1}}}

\newcommand{\Zn}{\ensuremath{Z^{(n)}}}

\newcommand{\Gain}[1]{\ensuremath{\mathcal{G}_{#1}}}

\newcommand{\eg}{\emph{e.g.,}\xspace}
\newcommand{\ie}{\emph{i.e.,}\xspace}

\submitted {10 April 2013}
\revised {23 May 2013}
\accepted{23 June 2013 }

\begin{document}

\title[Using RTDs for the Parallel Speedup Prediction of SAT Local Search]{Using Sequential Runtime Distributions for the Parallel Speedup Prediction of SAT Local Search}

\author[Alejandro Arbelaez,  Charlotte Truchet,  Philippe Codognet]{Alejandro Arbelaez\thanks{The first author was supported by the Japan Society for the Promotion of Science (JSPS) under the JSPS Postdoctoral Program and the \textit{kakenhi} Grant-in-aid for Scientific Research.}$^{1}$,  Charlotte Truchet$^2$,  Philippe Codognet$^{3}$ \\ $^{1}$JFLI / Univesity of Tokyo \\ $^{3}$JFLI-CNRS / UPMC / University of Tokyo \\ \email{\{arbelaez,codognet\}@is.s.u-tokyo.ac.jp} \\  $^{2}$LINA, UMR 6241/ University of Nantes  \\ \email{charlotte.truchet@univ-nantes.fr} }
\maketitle

\begin{abstract}
This paper presents a detailed analysis of the scalability and parallelization of local search algorithms for the Satisfiability problem. We propose a framework to estimate the parallel performance of a given algorithm by analyzing the runtime behavior of its sequential version. Indeed, by approximating the runtime distribution of the sequential process with statistical methods, the runtime behavior of the parallel process can be predicted by a model based on order statistics. We apply this approach to study the parallel performance of two SAT local search solvers, namely Sparrow and CCASAT, and compare the predicted performances to the results of an actual experimentation on parallel hardware up to 384 cores. We show that the model is accurate and predicts performance close to the empirical data. Moreover, as we study different types of instances (random and crafted), we observe that the local search solvers exhibit different behaviors and that their runtime distributions can be approximated by two types of distributions: exponential (shifted and non-shifted) and lognormal.
\end{abstract}

\section{Introduction}

Nowadays, SAT solvers
are very effective to solve problems in a wide variety of domains ranging from software verification to computational biology and automated planning. Broadly speaking, there are two main categories of SAT solvers: complete and incomplete. Complete solvers combine tree-based search with unit propagation, conflict-clause learning, and intelligent backtracking.
Incomplete solvers start with an initial assignment  for the variables (usually random); then the solver iteratively moves in the search space until a given stopping criteria is met. These solvers are very good at tackling large and difficult (random) instances.

Research on parallel SAT solvers have been rapidly increasing in the last decade, thanks to the the development and increasing availability of parallel hardware, such as multi-core architectures, GPGPUs, grids, cloud systems, and massively parallel supercomputers.
A well-known approach for parallel SAT solving is search-space splitting; it consists in dividing the problem space into several sub-spaces and exploring them in parallel. Another approach consists in building a parallel portfolio solver in which several algorithms compete and cooperate to solve a given problem instance. Motivated by the results of the recent SAT competitions, most researchers currently focus their attention on the development of parallel portfolios for multi-core architectures.
The computational benefit of the parallel portfolio is observed in both \textit{capacity solving} and speedup factor. \textit{Capacity solving}, or \textit{Solution Count Ranking}~\cite{VanGelder11}, refers to the ability of improving the total number of solved instances within a given timelimit, while speedup refers to the ability of reducing the runtime (w.r.t. the sequential solver) to solve individual instances. Previous work has been mainly focused on studying the \textit{capacity solving} of complete parallel SAT solvers, see \cite{SAT-survey} for a recent survey.

Up to now, most parallel SAT solvers have been designed for multi-core machines or small clusters with a few tens of processors. A key question is therefore to know if these approaches could scale up to massively parallel systems,
\ie with thousands or tens of thousands of cores.
To investigate this exiting new field of endeavor, we studied in this paper the parallel performance of several SAT solvers up to several hundreds of cores. Moreover, we propose a probabilistic model to estimate the parallel performance of local search algorithms for SAT, using a simple scheme for parallelization. By analyzing the sequential runtime, we can predict the parallel behavior and quantify the expected parallel speedup. More precisely, we first approximate the empirical sequential runtime distribution by a well-known statistical distribution (e.g. exponential or lognormal) and then derive the runtime distribution of the parallel version of the solver. Our model is related  to {\it order statistics}, a rather new domain of statistics~\cite{david2003order},  which is  the statistics  of sorted random draws.
This makes it possible to predict the parallel runtime of a given algorithm for any number of cores.

The main contributions of this paper are as follows. First, we present the application of a statistical model to predict and evaluate the performance of parallel local search algorithms for SAT. Moreover, extensive experimental results (up to 384 cores)  using state-of-the-art local search solvers showed that the predicted execution  times and speedups accurately match the empirical data and performance.
Second, we provide an understanding of the different speedups of parallel algorithms for SAT from a theoretical and empirical point of view for two different families of benchmarks.

This paper is organized as follows. After a brief presentation of parallel local search for SAT in Section \ref{section:parallel-sat}, Section \ref{section:runtime-distribution} describes the framework of runtime distributions and formally defines the probabilistic model used to predict the parallel performance of local search algorithms. Section \ref{section:experimental-settings} details extensive experimental results performed to evaluate the model.
Section \ref{section:conclusions} presents concluding remarks and future research directions.

\section{Parallel Local Search for SAT}
\label{section:parallel-sat}
Parallel         implementation         of        local         search
methods for combinatorial problems has been studied since the early 1990s,
when parallel machines started to become widely available~\cite{Pardalos95,Aarts95}.
Apart from domain-decomposition methods and population-based method (\eg
genetic algorithms),~\cite{Aarts95}  distinguishes between single-walk
and multi-walk methods for  Local Search.  Single-walk methods consist
in  using  parallelism  inside   a  single  search  process,  \eg  for
parallelizing   the   exploration  of   the   neighborhood.
Multi-walk   methods  (parallel   execution   of multi-start methods)
consist  in developing concurrent explorations of the  search space,
either  independently or cooperatively with some communication between processes.

It is now currently admitted that an easy and effective manner to parallelize local search solvers consists in executing in parallel multiple copies of a given solvers with or without cooperation. The non-cooperative approach has been used in the past to solve SAT and MaxSAT  instances. gNovelty+ \cite{gnovelty+}
executes multiple copies of gNovelty without cooperation until a solution is obtained or a given timeout is reached; and \cite{parallel-maxsat96} executes multiple copies of GRASP until an assignment which satisfies a given number of clauses is obtained.
Strategies to exploit cooperation between parallel SAT local search solvers have been studied in \cite{arbelaez-lion11} in the context of multi-core architectures with shared memory and in \cite{arbelaez-ictai12} in  massively parallel systems with distributed memory.

The analysis proposed in this paper for predicting performance on massively parallel systems is set in  the framework of independent multi-walk parallelism, as it seems to be the most promising way to deal with large-scale parallelism.
Cooperative algorithms might perform well on shared-memory machines with a few tens of cores,
but are difficult to extend efficiently to distributed hardware.

\section{Analysis using runtime distributions}
\label{section:runtime-distribution}
Most papers on the performance of stochastic local search algorithms
focus on the average execution time in order to
measure the performance of both sequential and parallel executions.
However, a more detailed analysis of the runtime behavior
could be done by looking at the execution time of the algorithm
(\eg cpu-time or number of iterations) as a random variable and
performing a statistical analysis of its probability distribution.

\subsection{Approximating Runtime Behaviors}

The notion of  {\it runtime distribution} has been introduced
by~\cite{Hoos98} to characterize the cumulative distribution function
of the execution time of stochastic algorithms.
Indeed, Stochastic Local Search~\cite{Hoos-book} can be considered in the
larger framework of {\it Las Vegas algorithms}, introduced a few
decades ago  by~\cite{Babai79}, i.e. randomized algorithms whose runtime
might vary  from one execution to another,  even with the same input.
It has been applied to study random 3-SAT problems with the Walk-SAT solver~\cite{Hoos-rtd},
combinatorial optimization problems with the GRASP metaheuristics~\cite{Aiex2002}
and path-planning problems with state-graph search algorithms (\eg A*)~\cite{Munoz12}.
The study of the runtime behavior  of parallel extensions of Las Vegas algorithm
in the framework of (independent) multi-walk processes has been proposed by~\cite{Predicting13},
which presents a model for predicting the parallel performance of a given Las Vegas algorithm
by the statistical analysis of its sequential version.
The runtime distribution has also been used to define optimal restart strategies in sequential and parallel algorithms in~\cite{Shylo11} and to provide bounds on the parallel expectation by \cite{luby_restarts}. However, in this paper we are using it to predict the parallel speedup in a multi-walk scheme from the study of the initial sequential problem distribution.

Indeed, since~\cite{Aarts95,Verhoeven96}, it is believed that combinatorial problems
can enjoy a linear speedup when implemented in parallel
by independent multi-walks. However, this has been proven only under the assumption
that the probability of finding a solution in a given time $t$
follows an exponential law, that is,
if the runtime behavior follows a (non-shifted) exponential distribution.
This behavior has been conjectured for SAT local search solvers
in~\cite{Hoos-rtd}, and confirmed experimentally
for the GRASP metaheuristics solver on some other classical combinatorial
problems~\cite{Aiex2002}.
The latter authors have also developed, in the context of combinatorial optimization,
a simple tool (\emph{tttplot}) to study the adequation
of a given runtime behavior with an exponential distribution~\cite{tttplots2007}.
The classical explanation for an exponential runtime behavior is
the fact that the solutions are uniformly distributed in the search space,
(and not regrouped in solution clusters~\cite{Maneva08}) and that the
random search algorithm is able to sample the search space in a uniform manner.
However,~\cite{Predicting13} shows that the
runtime distribution of local search solvers for combinatorial problems
can be not only exponential but also sometimes lognormal or shifted exponential,
in which cases the parallel speedup cannot be linear
and is asymptotically bounded.
Indeed, not all combinatorial problems show a perfect exponential behavior,
and we will see in this paper how this applies to SAT.

\subsection{\label{Min-Distribution} Min Distribution and Parallel Speed-up}

A general statistical model for studying the performance of Las Vegas algorithms
and predicting the parallel performance of their parallel multi-walk extensions has been
recently proposed in~\cite{Predicting13}.
We will now present a brief summary of this model,
which will be used in the rest of the paper to study the behavior
of two local search solvers on a variety of SAT instances.

Let $Y$ be the runtime of a Local Search algorithm on a  given problem instance.
It can be considered as a random variable
with values in \Nat~({\it number of iterations}), or in \Rat~({\it cpu-time}). In general, it is more convenient to consider distributions  with values in
\Reals~because  calculations are  easier.
$Y$ can  be studied  through its
cumulative  distribution,   which  is  by   definition,  the  function
\cumulative{Y}  s.t.  $\cumulative{Y}(x)=\Prob{Y \leq  x}$.
By definition, the distribution of $Y$ is the derivative of \cumulative{Y}:
$\Distrib{Y}=\cumulative{Y}'$.
 The  expectation of  $Y$ is
defined as $\Esp{Y}=\int_0^\infty t \Distrib{Y}(t) dt$

Assume that $n$ copies of the base algorithm  are running in parallel on $n$ cores.
The first process finding a solution kills
all  others,   and  the  overall parallel algorithm  terminates.   The  $i$-th  process
corresponds  to   a  draw  of  a  random   variable  $X_i$,  following
the distribution $\Distrib{Y}$.  The variables $X_i$ are thus
independent and  identically distributed (i.i.d.).   The computation
time of the whole parallel process is also a random variable, let \Zn,
with  a distribution  \Distrib{\Zn} that  depends both  on $n$  and on \Distrib{Y}.
 Since   all  the  $X_i$  are   i.i.d.,  the  cumulative
distribution $\cumulative{\Zn}$  and the distribution $\Distrib{\Zn}$ can be computed as follows:

\[
\begin{array}{ccccc}
\centering
\cumulative{\Zn}  &=&  \Prob{\Zn \leq x}
  &=&  ~~\Prob{\exists i \in \{1...n\}, X_i \leq x}\\
& = & ~~1 - \Prob{\forall i \in \{1...n\}, X_i > x}~~
& = & 1 - \left( 1 - \cumulative{Y}(x)\right)^n \\
\Distrib{\Zn}&=& \multicolumn{3}{l}{ \left(1 -( 1 - \cumulative{Y}\right)^n) '
 = n\Distrib{Y} (1-\cumulative{Y})^{n-1} } \\
\end{array}
\]

Thus, knowing  the distribution  for the base  algorithm $Y$,  one can
calculate the distribution for $\Zn$.
The formula  shows that the  parallel algorithm favors short  runs, by
killing slower processes. Thus, compared to the distribution of $Y$,
the distribution of $\Zn$ moves toward the origin and is more peaked.

We can also compute the expectation $\Esp{\Zn}$ for the parallel process,
from which we derive the expected speed-up \Gain{n} of the parallel algorithm
versus the sequential one:
\begin{eqnarray*}
\Esp{\Zn}
&=& n \int_0^\infty t \Distrib{Y}(t) (1-\cumulative{Y}(t))^{n-1}dt \\
\Gain{n}&=&\Esp{Y}/\Esp{\Zn}
\end{eqnarray*}
Again, no explicit general formula can be  computed and the  expression of the
speed-up will depend on the distribution of $Y$.
We   will  thus study in   the  following   different  specific
distributions. This computation of  the speed-up is actually
related to a field of statistics called
order   statistics,   see   \cite{david2003order}   for   a   detailed
presentation.  Order statistics are the statistics of sorted random draws.
For  instance, the  first order  statistics of  a distribution  is its
minimal   value.
For predicting  the speedup, we  are indeed
interested  in computing the  expectation of  the distribution  of the
minimum  draw. As the  above  formula  suggests,  this may  lead  to
heavy    calculations,     but    recent    studies     such    as
\cite{Nadarajah2008}  give  explicit formulas  defining  this quantity  for
several classical probability distributions.

\subsection{Exponential and Lognormal Distributions}
\label{exponential-distribution}
\label{lognormal-disitribution}

Assume that $Y$ has a shifted exponential distribution,
as would be the case for an ideal randomized algorithm.
The minimum distribution can be computed by a direct integration:
$\Distrib{Y}(t)$=$\lambda e^{-\lambda (t-x_0)}$ for t$>$ 0; $\Esp{Y}$=$x_0+1/\lambda$; and $\Distrib{\Zn}(t)$=$n$$\lambda e^{-n \lambda (t-x_0)}$ for $t>0$.

In case of a non-shifted exponential,  $x_0=0$ and
the  speed-up is thus equal to the number of cores $n$,
up to infinity.
This  case has  already  been studied by~\cite{Aarts95}.
However for  $x_0>0$, the  speed-up admits a  finite limit,
even when $n$ tends to infinity,
which  is $\frac{x_0+\frac{1}{\lambda}}{x_0}=1+ \frac{1}{x_0\lambda}$.
The closer to zero $x_0$ is, the higher the limit.

Other distributions  can be considered,  depending on the  behavior of
the  base   algorithm.  We  will   study  the  case  of   a  lognormal
distribution, which is the log  of a gaussian distribution, because it
will appear in the following experiments for some instances.
The  lognormal distribution
has  two  parameters,  the  mean  $\mu$  and  the  standard  deviation
$\sigma$.
Formally,
a (non-shifted) lognormal distribution is defined as:
$ \Distrib{Y}(t)=
\frac{e^{-\frac{(-\mu+log(t))^2}{2\sigma^2}}}{\sqrt{2\pi} (t) \sigma}$

The  formulas for the  distribution of  \Zn, its
expectation  and  the   theoretical  speed-up  are  quite  complicated
to compute, but \cite{Nadarajah2008} gives  an explicit formula for all
the  moments  of lognormal  order  statistics  with  only a  numerical
integration  step, from  which  we  can derive  a  computation of  the
speed-up.
As for the shifted exponential, it can be shown that
the speed-up curve of the lognormal distribution admits a finite limit.

\section{Experimental Settings and Results}
\label{section:experimental-settings}

This section describes the benchmark instances used for tests, and we focus our attention on two well-known problem families: random and crafted instances. Moreover, we consider the two best local search solvers from the previous SAT competition:
CCASAT \cite{ccasat} and Sparrow \cite{sparrow}. Both solvers were used with their default parameters and with a timeout of 3 hours for each experiments.
All the experiments were performed on the Grid'5000 platform, the French national grid for research. We used a 44-node cluster with 24 cores (2 AMD Opteron 6164 HE processors at 1.7 Ghz) and 44 GB of RAM per node. We experimented with 10 random instances (6 around the phase transition) and 10 crafted instances (see the appendix for a complete presentation of the instances).

In order to obtain the empirical data for the theoretical distribution
(predicted by our model from the sequential runtime distribution),
we performed 500 runs of the sequential algorithm. The Mathematica software \cite{mathematica}, version 8.0,
was used to estimate the parameters of the theoretical distributions
and to integrate numerically the formulas of the lognormal distribution.
In order to evaluate the accuracy of the learned statistical model, we performed 50 runs of the multi-walk parallel algorithms. The empirical speedup for a given parallel algorithm is calculated against the mean performance of its sequential version as follows:
$Speedup = \frac{\textit{Mean}(Solver~on~1\textit{ core})}{\textit{Mean}(Solver~on~\textit{N cores})} $

\subsection{Experimental results}
\label{section:experiments}
In this section, we start by presenting the empirical and estimated results for random and crafted instances; then we present a general analysis of the results.

\label{section:experiments-random}

We start our analysis with Table \ref{table:random-seq}, which presents initial statistics for the sequential version of Sparrow and CCASAT. We present the minimum, maximum, and mean runtime values, as well as the outcome of the Kolmogorov-Smirnov (KS) test for two types of distributions: shifted exponential and lognormal.
In the following tables, bold numbers indicate the distribution chosen to predict the performance of a given solver.

\begin{table}[h!tb]
\small
\centering
\begin{tabular}{ccccccc}
\hline
\hline
\multirow{2}{*}{Instance} & \multirow{2}{*}{Alg} & \multirow{2}{*}{Min} & \multirow{2}{*}{Max} & \multirow{2}{*}{Mean} & \multicolumn{2}{c}{p-value}  \\
\cline{6-7}
& & & & & Shifted exp. dist. & Lognormal dist.  \\
\hline
\multirow{2}{*}{Rand-1} & Sparrow &  98.0 & 4860.0 & 793.9 & 5.6$\cdot10^{-19}$ & \textbf{0.76}\\
 & CCASAT & 103.6 & 1340.1 & 458.5 & 8.4$\cdot10^{-55}$ &  \textbf{0.96} \\
\noalign{\vspace {.05cm}}
\multirow{2}{*}{Rand-2} & Sparrow & 91.8 & 5447.0 & 1007.5 & 1.8$\cdot10^{-20}$  & \textbf{0.91} \\
 & CCASAT & 108.8 & 1652.9 & 497.9 & 5.1$\cdot10^{-58}$ &  \textbf{0.71}  \\
\noalign{\vspace {.05cm}}
\multirow{2}{*}{Rand-3} & Sparrow & 104.8 & 3693.7 & 797.0 & 2.9$\cdot10^{-26}$ & \textbf{0.64} \\
 & CCASAT & 126.48 & 1125.4 & 359.6 & 2.3$\cdot10^{-108}$ & \textbf{0.90} \\
 \noalign{\vspace {.05cm}}
\multirow{2}{*}{Rand-4} & Sparrow &  162.4 & 3037.8 & 781.5 & 1.6$\cdot10^{-38}$ & \textbf{0.03} \\
 & CCASAT & 132.5 & 980.9 & 382.0 & 1.5$\cdot10^{-117}$ &  \textbf{0.92} \\
\noalign{\vspace {.05cm}}
\multirow{2}{*}{Rand-5} & Sparrow &  164.0 & 7946.3 & 952.4 & 1.8$\cdot10^{-31}$ &  \textbf{0.16} \\
 & CCASAT & 158.6 & 1177.9 & 403.1 & 5.1$\cdot10^{-134}$ &  \textbf{0.20} \\
\noalign{\vspace {.05cm}}
\multirow{2}{*}{Rand-6} & Sparrow &  142.0 & 4955.8 & 763.5 & 1.6$\cdot10^{-31}$ & \textbf{0.64} \\
 & CCASAT & 142.9 & 890.9 & 354.1 & 9.6$\cdot10^{-137}$ &  \textbf{0.46} \\
\noalign{\vspace {.05cm}}
\hline
\multirow{2}{*}{Rand-7} & Sparrow &  35.5 & 10637.4 & 3464.2 & \textbf{0.01} & 1.5$\cdot10^{-4}$ \\ 
 & CCASAT & 61.6 & 6419.1 & 1801.0 & 1.0$\cdot10^{-5}$  &  \textbf{0.13} \\
\noalign{\vspace {.05cm}}
\multirow{2}{*}{Rand-8} & Sparrow & 23.2 & 10738.0 & 3412.9 & \textbf{0.05} &  4.7$\cdot10^{-4}$\\ 
 & CCASAT &  35.9 & 10443.7 & 2007.6 & \textbf{0.50}  & 0.03 \\
\noalign{\vspace {.05cm}}
\multirow{2}{*}{Rand-9} & Sparrow &   6.8 & 5935.8 & 1028.2 & \textbf{0.23} & 3.7$\cdot10^{-3}$ \\ 
 & CCASAT &  18.1 & 2830.4 & 476.8 & 7.0$\cdot10^{-3}$ & \textbf{0.03} \\
\noalign{\vspace {.05cm}}
\multirow{2}{*}{Rand-10} & Sparrow &  19.0 & 10800.0 & 1726.3 &  \textbf{0.65} & 0.15 \\
 & CCASAT & 19.8 & 4854.5 & 758.4 & 7.6$\cdot10^{-10}$ & \textbf{0.18} \\
\hline
\hline
\end{tabular}
\caption{Performance of sequential algorithms on random instances}
\label{table:random-seq}
\end{table}

The KS test compares a set of empirical measures to a given theoretical distribution. Its outcome is a {\it p-value}, indicating how likely it is that the measures admits the theoretical distribution. The classical threshold for the p-value is 0.05. For greater p-values, the KS test succeeds (more precisely, the null hypothesis is not rejected),
and the empirical distribution can be approximated by the theoretical one with good confidence.

The results presented in this table are consistent with the results of the previous SAT competition (random category) where CCASAT greatly outperformed Sparrow.  For this set of instances, we choose the shifted exponential distribution in lieu of the exponential distribution as the Min runtime value for the reference solvers is not negligible compared to its mean value across 500 executions (about 100 times smaller in the best case).

As can be seen from the table, both solvers report a tendency which indicates that the empirical data for instances around the phase transition  are better approximated by a lognormal distribution; all these instances pass the KS test with a confidence level (p-value) above 0.05, except for Sparrow on rand-4.

For instances outside the phase transition, Sparrow reports enough statistical evidence to infer that the shifted exponential distribution fits better the empirical data. For CCASAT, 3 out of 4 instances outside the phase transition are better characterized with a lognormal distribution and the remaining instance pass the KS test for the shifted exponential distribution.

\begin{table}[h!]
\small
\centering
\begin{tabular}{cccccccccc}
\hline
\hline
\multirow{2}{*}{Instance} & \multirow{2}{*}{ } & \multicolumn{4}{c}{Sparrow - Runtime on $k$ cores}    & \multicolumn{4}{c}{CCASAT - Runtime on $k$ cores} \\
& & 48 & 96 & 192 & 384 &  48 & 96 & 192 & 384 \\
\hline
\multirow{2}{*}{Rand-1} & Actual  &  163.8 & 140.4 & 125.2 & 113.7 & 160.0 & 143.0 & 122.8 & 112.0 \\
& Predicted & 133.8 & 110.5  & 92.7 & 78.8 & 137.7 & 120.6 & 106.7 & 95.3 \\
\noalign{\vspace {.05cm}}
\multirow{2}{*}{Rand-2} & Actual & 213.2 & 191.4 & 166.2 & 142.5 & 186.8 & 169.3 & 159.3 & 142.8\\
& Predicted & 183.5 & 152.8 & 129.2 & 110.6 & 153.4 & 134.7 & 119.6 & 107.1\\
\noalign{\vspace {.05cm}}
\multirow{2}{*}{Rand-3}  & Actual & 175.9 & 151.2 & 135.8 & 123.5 & 166.7 & 155.6 &143.5 &132.2 \\
& Predicted & 183.5 & 152.8 &129.2 &110.6 & 153.4 &134.7 & 119.6 & 107.1 \\ \noalign{\vspace {.05cm}}
\multirow{2}{*}{Rand-4} & Actual  & 202.3 & 179.2 & 159.5 & 141.8 & 193.1 & 176.0 & 169.4 & 158.7 \\
& Predicted & 175.7 & 149.5 & 128.9 & 112.4 & 170.6 & 155.9 & 143.5 &132.8 \\
\noalign{\vspace {.05cm}}
\multirow{2}{*}{Rand-5} & Actual & 219.6 & 201.0 & 182.5 & 161.9 & 212.2 & 191.3 & 176.8 & 165.8 \\
& Predicted & 185.0 & 155.3 & 132.3 & 114.0 & 179.8 & 164.3 & 151.2 & 140.0 \\
\noalign{\vspace {.05cm}}
\multirow{2}{*}{Rand-6} & Actual & 185.5 & 167.1 & 150.3 & 137.5 & 190.9 & 179.3 & 168.4 & 153.4 \\
& Predicted & 158.3 &133.6 & 114.4 & 99.1 & 160.6 & 147.0 & 135.4 & 125.6 \\
\noalign{\vspace {.05cm}}
\hline
\multirow{2}{*}{Rand-7} & Actual & 151.2 & 102.7 & 63.8 & 51.1 & 22.9 & 33.7 & 54.3 & 67.8 \\
& Predicted & 195.8 & 143.0 & 107.3 & 82.3  & 182.8 & 142.6 & 113.7 & 92.2 \\
\noalign{\vspace {.05cm}}
\multirow{2}{*}{Rand-8} & Actual & 126.6 &  81.9 & 51.1 & 30.9 & 131.8 &  83.9 & 64.8 & 39.7 \\
& Predicted & 93.8 & 58.5 & 40.8 & 32.0 & 76.9 & 56.4 & 46.1 & 41.0 \\
\noalign{\vspace {.05cm}}
\multirow{2}{*}{Rand-9} & Actual & 33.9 & 18.4 & 13.1 & 9.0 & 45.0 & 31.0 & 22.7 & 16.3 \\
& Predicted &  28.1 & 17.4 & 12.1 & 9.4 & 38.5 & 29.4 & 23.0 & 18.3 \\
\noalign{\vspace {.05cm}}
\multirow{2}{*}{Rand-10} & Actual & 63.4 & 48.9 & 40.7 & 30.9 & 113.8 & 94.7 & 72.9 & 54.2 \\
& Predicted &  54.6 & 36.8 & 27.9 & 23.4 & 105.6 & 85.3 & 70.2 &  58.6 \\
\hline
\hline
\end{tabular}
\caption{Runtimes for random instances up to 384 cores}
\label{table:random-parallel-runtime}
\end{table}

Let's now look at the parallel performance of the solvers. Table \ref{table:random-parallel-runtime} (resp. Table \ref{table:random-parallel-speedup}) shows the empirical and predicted runtime (resp. speedup)
for both Sparrow and CCASAT on all instances using 48, 96, 192, and 384 cores.
In Table~\ref{table:random-parallel-speedup}, we observe an important difference in the speedup factor
 between the two solvers which suggest that in general Sparrow scales better than CCASAT.

\begin{table}[h!]
\small
\centering
\begin{tabular}{cccccccccc}
\hline
\hline
\multirow{2}{*}{Instance} & \multirow{2}{*}{ } & \multicolumn{4}{c}{Sparrow - Speedup on $k$ cores}    & \multicolumn{4}{c}{CCASAT - Speedup on $k$ cores} \\
& & 48 & 96 & 192 & 384 &  48 & 96 & 192 & 384 \\
\hline
\multirow{2}{*}{Rand-1} & Actual  &  4.8 & 5.6 & 6.3 & 6.9 & 2.8 & 3.2 & 3.7 & 4.0 \\
& Predicted & 5.9 & 7.1 & 8.5 & 10.0 & 3.3 & 3.8 & 4.3 & 4.8 \\
\noalign{\vspace {.05cm}}
\multirow{2}{*}{Rand-2} & Actual & 4.7 & 5.2 & 6.0 &7.0 & 2.6 & 2.9 & 3.1 & 3.4 \\
& Predicted & 5.4 & 6.5 & 7.7 & 9.0 & 3.2 & 3.6 & 4.1 & 4.6 \\
\noalign{\vspace {.05cm}}
\multirow{2}{*}{Rand-3} & Actual & 4.5 & 5.2 & 5.8 & 6.4 & 2.1 & 2.3 & 2.5 & 2.7 \\
& Predicted & 5.4 & 6.5 & 7.7 & 9.0 &  3.2 & 3.6 & 4.1 & 4.6 \\
\noalign{\vspace {.05cm}}
\multirow{2}{*}{Rand-4} & Actual & 3.8 & 4.3 & 4.8 & 5.5  & 1.9 & 2.1 & 2.2 & 2.4 \\
& Predicted & 4.4 & 5.1 & 6.0 & 6.9 & 2.2 & 2.4 & 2.6 & 2.8 \\
\noalign{\vspace {.05cm}}
\multirow{2}{*}{Rand-5} & Actual &  4.3 & 4.7 & 5.2 & 5.8 & 1.9 &  2.1 & 2.2 & 2.4 \\
& Predicted & 5.0 & 6.0 & 7.0 & 8.1 & 2.2 & 2.4 & 2.6 & 2.8  \\
\noalign{\vspace {.05cm}}
\multirow{2}{*}{Rand-6} & Actual &  4.1 & 4.5 & 5.0 & 5.5 &  1.8 & 1.9 & 2.1 & 2.3 \\
& Predicted & 4.7 & 5.6 & 6.6 & 7.6 & 2.2 & 2.4 & 2.6 & 2.8  \\
\noalign{\vspace {.05cm}}
\hline
\multirow{2}{*}{Rand-7} & Actual & 22.9 & 33.7 & 54.3 & 67.8 & 9.5 & 13.7 & 18.5 & 22.7 \\
& Predicted & 32.3 & 48.5 & 64.8 & 77.8  & 10.5 & 13.4 &  16.9 & 20.8 \\
\noalign{\vspace {.05cm}}
\multirow{2}{*}{Rand-8} & Actual & 26.9 & 41.7 & 66.8 & 110.6 & 15.2 & 23.9 & 30.9 & 50.5 \\
& Predicted &  36.3 & 58.3 & 83.5 & 106.5 & 26.0 & 35.5 & 43.4 & 48.9 \\
\noalign{\vspace {.05cm}}
\multirow{2}{*}{Rand-9} & Actual & 30.3 & 55.8 & 78.1 & 114.2 & 10.5 & 15.3 & 20.9 &  29.1 \\
& Predicted &  36.5 & 58.8 & 84.6 &  108.3 & 13.2 & 17.3 & 22.2 &  27.9 \\
\noalign{\vspace {.05cm}}
\multirow{2}{*}{Rand-10} & Actual & 27.2 & 35.2 & 42.3 & 55.7 & 6.6 & 8.0 & 10.3 & 13.9 \\
& Predicted & 31.6 & 46.8 & 61.7 & 73.4  & 7.3 & 9.1 & 11.1 & 13.3 \\
\hline
\hline
\end{tabular}
\caption{Speedup for random instances up to 384 cores}
\label{table:random-parallel-speedup}
\end{table}

\begin{figure}[h!]
\centering
	\subfigure[Empirical CDF vs. theoretical CDF (rand-4)]{\includegraphics[scale=0.3]{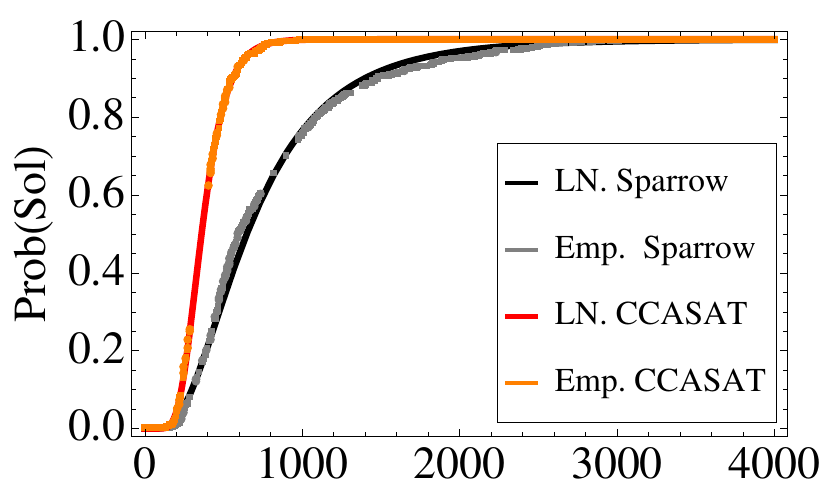}\label{fig:random-3-CDF}}\hspace{1mm}
	\subfigure[Empirical CDF vs. predicted CDF using 384 cores (rand-4)]{\includegraphics[scale=0.3]{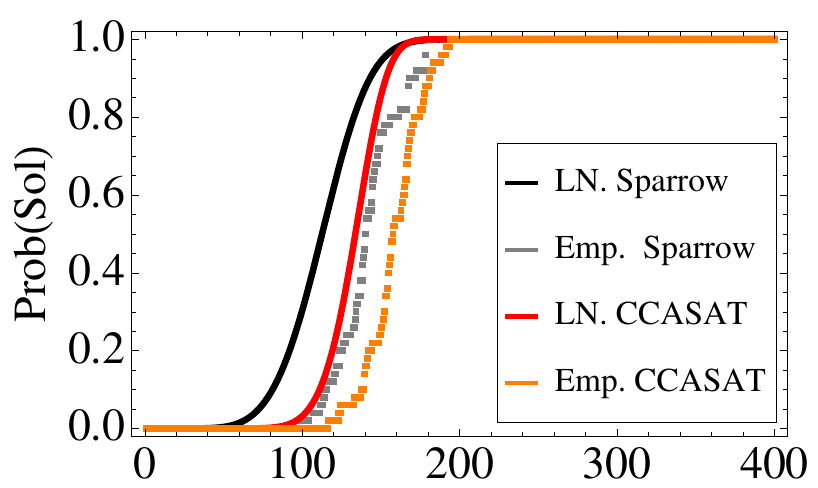}\label{fig:random-3-CDF-384cores}}\hspace{1mm}
	\subfigure[Speedup (rand-4)]{\includegraphics[scale=0.3]{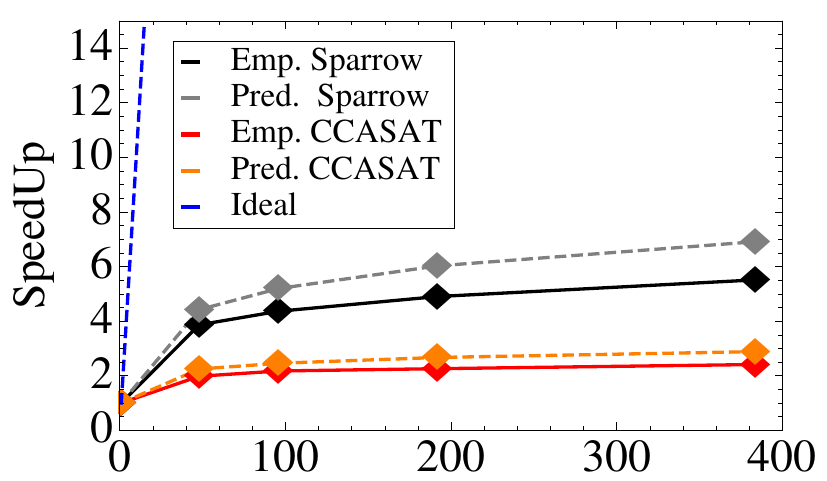}\label{fig:random-3-speedup}}
	\subfigure[Empirical CDF vs. theoretical CDF (rand-7)]{\includegraphics[scale=0.3]{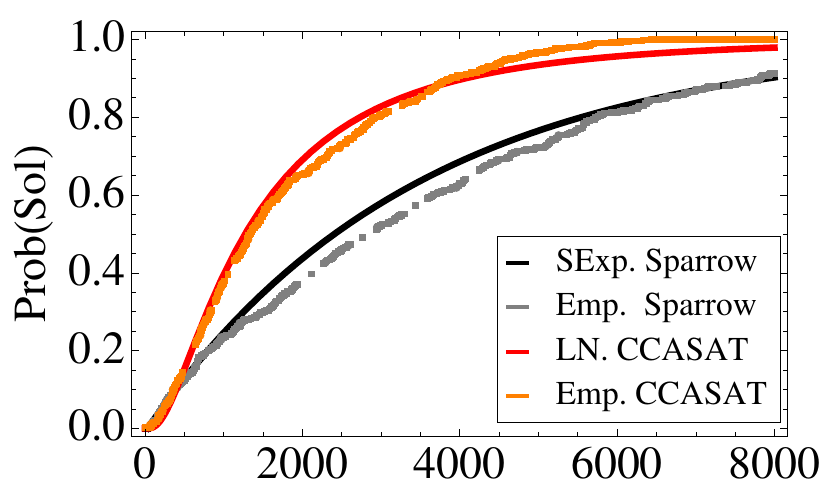}\label{fig:random-7-CDF}}
	\subfigure[Empirical CDF vs. predicted CDF using 384 cores (rand-7)]{\includegraphics[scale=0.3]{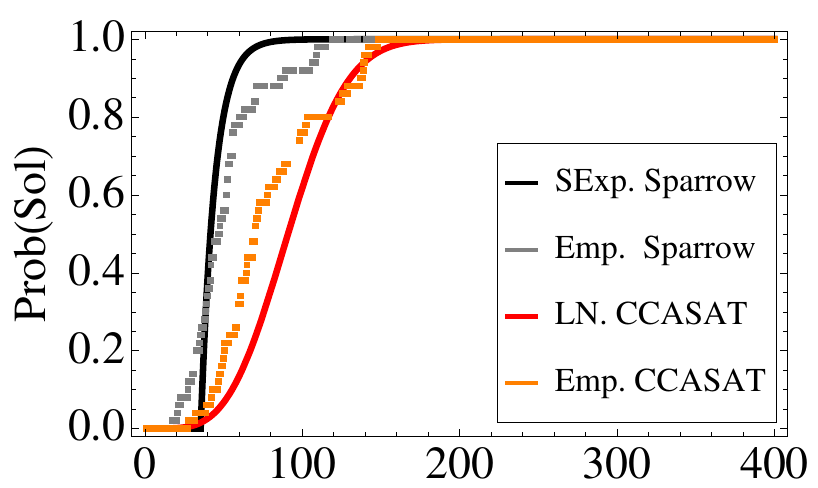}\label{fig:random-7-CDF-384cores}}
	\subfigure[Speedup (rand-7)]{\includegraphics[scale=0.3]{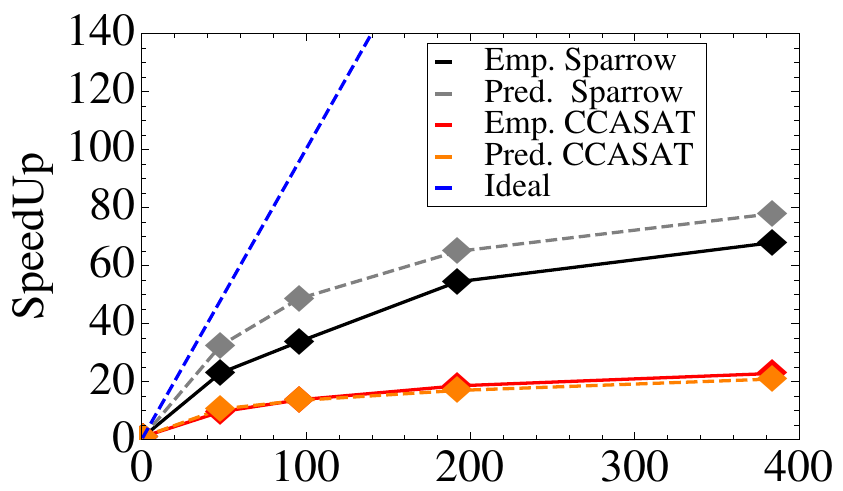}\label{fig:random-7-speedup}}
\caption{Performance summary to solve rand-4 and rand-7}
\label{fig:random-performance}
\end{figure}

Figure \ref{fig:random-performance} shows a performance summary of the reference solvers to tackle an instance on the phase transition (rand-4) and another instance outside the phase transition (rand-7). The \textit{y-axis} gives the probability ($\Prob{Y \leq  x}$) of finding a solution in a time less or equal to $x$ and the \textit{x-axis} gives the runtime in seconds.
From now on, in all figures `Emp' stands for Empirical distribution, `LN' stands for lognormal distribution,
and  `SExp' stands for shifted exponential distribution.
As expected CCASAT dominates the performance on one core. For example to solve rand-4, CCASAT reports $\Prob{\textit{Y} \leq \textit{16-mins}}$ $\approx$ 1.0, while Sparrow reports $\Prob{\textit{Y} \leq \textit{16-mins}}$ $\approx$ 0.75.
Figures \ref{fig:random-3-speedup} and \ref{fig:random-7-speedup} show that for CCASAT increasing the number of cores does not significantly improve the solving time. Consequently, Sparrow becomes more effective for a large number of cores. Therefore, Figures \ref{fig:random-3-CDF-384cores} and \ref{fig:random-7-CDF-384cores} show that Sparrow is  better than CCASAT when using 384 cores.
Interestingly, the same pattern is observed for other random instances (see Table \ref{table:random-parallel-runtime}).

To illustrate the power of the predicted model, in Figure \ref{fig:random-3-speedup} we present the predicted and empirical speedup curves for CCASAT and Sparrow.
Here it can be observed that in both cases the predicted curve follows the same shape as the empirical one. Moreover, It is also important to note that the speedup factor of the reference solvers for this problem family is far from linear (ideal), a phenomenon described by the predicted model.

Finally, it can also be observed that random instances around the phase transition exhibit a lower speedup factor than the remaining random instances. For instance, the best empirical speedup factor obtained for instances in the phase transition is  7.0 for Sparrow  and 3.4 for CCASAT; and the best speedup factor obtained for instances outside the phase transition is 114.2 for Sparrow and 50.5 for CCASAT.

\begin{figure}[h!]
\centering
	\subfigure[Empirical CDF vs. theoretical CDDs (CCASAT)]{\label{figure:crafted-0-ccasat}\includegraphics[scale=0.3]{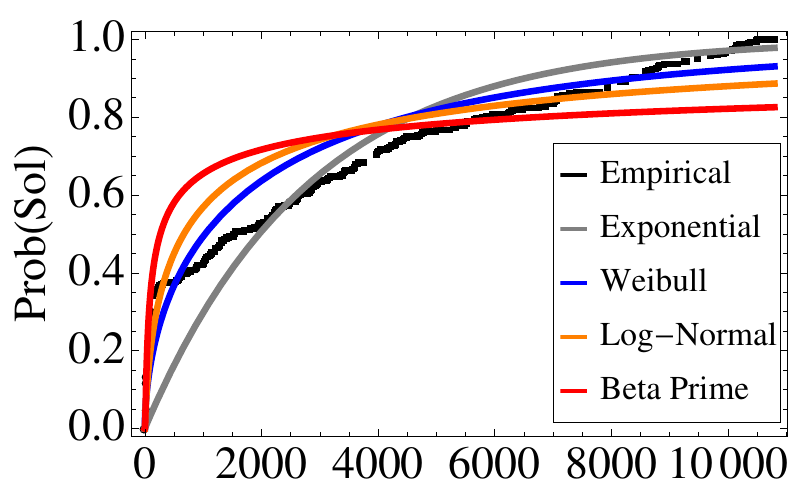}}\hspace{20mm}
	\subfigure[Empirical CDF vs. theoretical CDF (Sparrow)]{\label{figure:crafted-0-sparrow-cdf}\includegraphics[scale=0.3]{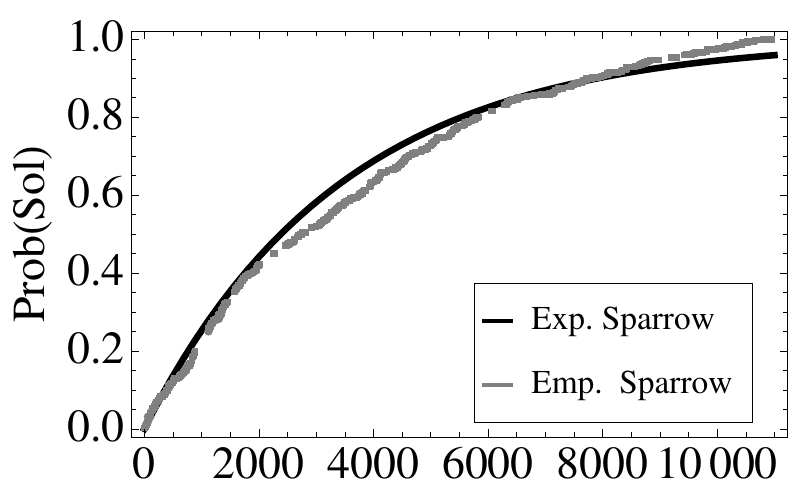}}
	\subfigure[Empirical CDF vs. predicted CDF using 384 cores (Sparrow)]{\label{figure:crafted-0-sparrow-cdf-384cores}\includegraphics[scale=0.3]{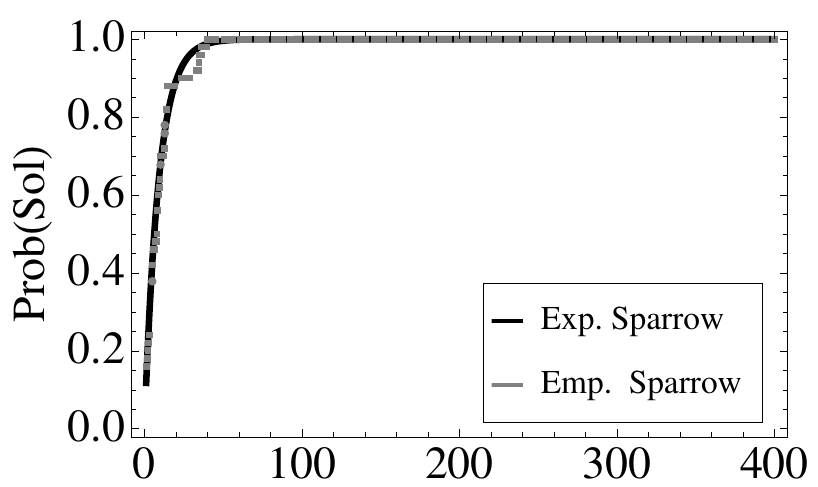}}\hspace{20mm}
	 \subfigure[Speedup (Sparrow)]{\label{figure:crafted-0-sparrow-speedup}\includegraphics[scale=0.3]{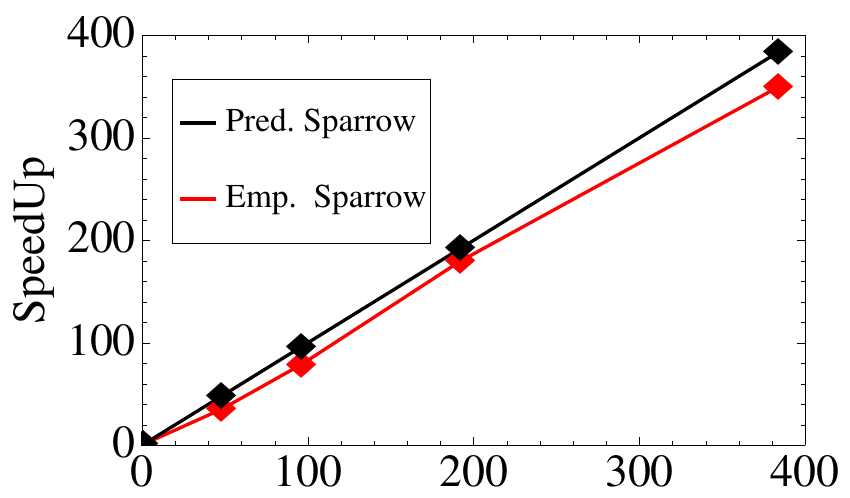}}
\caption{Performance summary on crafted-1}
\label{figure:craft-0-sparrow}
\end{figure}

Let's switch our attention now to crafted instances, for which we have to treat differently CCASAT and Sparrow.
For CCASAT, we were unable to find a theoretical distribution which fits the empirical data.
It should be also noticed that CCASAT has been mainly designed and tuned to handle random instances.
Let us look for instance at Figure \ref{figure:crafted-0-ccasat}, which depicts the cumulative runtime distribution of CCASAT to solve Crafted-1 using the two reference distributions detailed in this paper (lognormal and exponential) and two extra distributions (Weibull and beta-prime). None of the theoretical distributions seems to be a good approximation of the empirical data. More precisely, the KS test reported a p-value of 2.7$\cdot10^{-7}$ (lognormal); 7.0$\cdot10^{-24}$ (exponential); 2.4$\cdot10^{-6}$ (Weibull); and 6.9$\cdot10^{-15}$ (beta-prime).
Therefore, none of the theoretical distributions pass KS test with a high-enough p-value. We also experimented with other instances and observed a similar behavior.

\begin{table}[h!]
\small
\centering
\begin{tabular}{ccccccc}
\hline
\hline
\multirow{2}{*}{Instance} & \multirow{2}{*}{Alg} & \multirow{2}{*}{Min} & \multirow{2}{*}{Max} & \multirow{2}{*}{Mean} & \multicolumn{2}{c}{p-value}  \\
& & & & & Exp. dist. & Lognormal dist.  \\
\hline
Crafted-1 & Sparrow & 9.9 & 10800.0 & 3440.3 & \textbf{0.02} & 1.2$\cdot10^{-4}$ \\
\noalign{\vspace {.05cm}}
Crafted-2 & Sparrow & 1.0 & 10800.0 & 2711.2 & \textbf{0.57} & 1.4$\cdot10^{-4}$ \\
\noalign{\vspace {.05cm}}
Crafted-3 & Sparrow & 8.7 & 10800.0 & 3432.7 & \textbf{0.14} & 1.1$\cdot10^{-3}$ \\
\noalign{\vspace {.05cm}}
Crafted-4 & Sparrow & 2.2 & 10800.0 & 2701.6 & \textbf{0.11} & 9.6$\cdot10^{-3}$ \\
\noalign{\vspace {.05cm}}
Crafted-5 & Sparrow & 4.1 & 10800.0 & 1564.1 & \textbf{0.95} & 9.2$\cdot10^{-4}$ \\
\noalign{\vspace {.05cm}}
Crafted-6 & Sparrow & 2.9 & 10800.0 & 3599.6 & \textbf{0.01} & 1.0$\cdot10^{-5}$ \\
\noalign{\vspace {.05cm}}
Crafted-7 & Sparrow & 4.4 & 10800.0 & 3598.7 & \textbf{0.01} & 7.8$\cdot10^{-6}$ \\
\noalign{\vspace {.05cm}}
Crafted-8 & Sparrow & 3.5 & 5456.0 & 972.046 & \textbf{0.67} & 0.17 \\
\noalign{\vspace {.05cm}}
Crafted-9 & Sparrow & 1.9 & 7876.5 & 1298.24 & \textbf{0.97} & 7.0$\cdot10^{-3}$ \\
\hline
\hline
\end{tabular}
\caption{Sequential performance of Sparrow on crafted instances}
\label{table:sparrow-crafted}
\end{table}

For Sparrow on all crafted instances, the KS test shows a much better p-value for the exponential distribution than for the lognormal one, see Table \ref{table:sparrow-crafted}.
The confidence level is quite high for the instances Crafted-2,-3,-4,-5,-8,-9, with p-value up to 0.97,
 while the p-value is between 0.01 and 0.02 for Crafted-1,-6,-7.
 Also, as the minimum runtime is much smaller than the mean (at least 300 times smaller), we can approximate the empirical data by a non-shifted exponential distribution~\cite{Predicting13}.

As can be seen in Table \ref{table:sparrow-crafted-parallel} the multi-walk parallel approach scales well for Sparrow on crafted instances as the number of cores increases. Indeed a nearly linear speedup is obtained for nearly all the instances. As expected, the speedup predicted by our model is optimal, and this result is consistent with those obtained in \cite{Hoos-rtd}.

\begin{table}[h!]
\small
\centering
\begin{tabular}{cccccccccc}
\hline
\hline
\multirow{2}{*}{Instance} & \multirow{2}{*}{ } & \multicolumn{4}{c}{Runtime on $k$ cores}    & \multicolumn{4}{c}{Speedup ok $k$ cores} \\
& & 48 & 96 & 192 & 384 &  48 & 96 & 192 & 384 \\
\hline
\multirow{2}{*}{Crafted-1} & Actual & 97.7 & 43.7 & 19.1 & 9.8 &   35.1 & 78.6 & 179.6 & 349.8\\
& Predicted & 71.6 & 35.8 & 17.9 & 8.9 & 48.0 & 96.0 & 192.0 & 384.0 \\
\noalign{\vspace {.05cm}}
\multirow{2}{*}{Crafted-2} & Actual &  67.8 & 36.4 & 17.5 & 7.2  & 39.9 & 74.4 & 154.7 & 375.2 \\
& Predicted & 56.4 & 28.2 & 14.1 & 7.0 & 48.0 & 96.0 & 192.0 & 384.0 \\
\noalign{\vspace {.05cm}}
\multirow{2}{*}{Crafted-3} & Actual &  94.8 & 49.3 & 23.2 & 11.9 &  36.1 & 69.6 & 147.6 & 286.1  \\
& Predicted & 71.5 & 35.7 & 17.8 & 8.9 & 48.0 & 96.0 & 192.0 & 384.0  \\
\noalign{\vspace {.05cm}}
\multirow{2}{*}{Crafted-4} & Actual & 87.5 & 42.0 & 17.3 & 9.7 &   30.8 & 64.2 & 155.4 & 277.8 \\
& Predicted & 56.2 & 28.1 & 14.0 & 7.0 & 48.0 & 96.0 & 192.0 & 384.0  \\
\noalign{\vspace {.05cm}}
\multirow{2}{*}{Crafted-5} & Actual & 33.7 & 15.1 & 7.6 & 4.2 &  46.3 & 103.2 & 204.1 & 371.6 \\
& Predicted & 32.5 & 16.2 & 8.1 & 4.0 & 48.0 & 96.0 & 192.0 & 384.0  \\
\noalign{\vspace {.05cm}}
\multirow{2}{*}{Crafted-6} & Actual & 130.0 & 69.8 & 25.6 & 12.8 &  27.6 & 51.5 & 140.5 & 279.5 \\
& Predicted & 74.9 & 37.4 & 18.7 & 9.3  & 48.0 & 96.0 & 192.0 & 384.0   \\
\noalign{\vspace {.05cm}}
\multirow{2}{*}{Crafted-7} & Actual & 95.0 & 51.3 & 28.4 & 11.6 &  37.8 & 70.0 & 126.3 & 308.0 \\
& Predicted & 74.9 & 37.4 & 18.7 & 9.3 & 48.0 & 96.0 & 192.0 & 384.0    \\
\noalign{\vspace {.05cm}}
\multirow{2}{*}{Crafted-8} & Actual & 17.2 & 10.8 & 5.3 & 2.6 &  56.4 & 89.6 & 181.1 & 363.6 \\
& Predicted & 20.2 & 10.1 & 5.0 & 2.5 & 48.0 & 96.0 & 192.0 & 384.0  \\
\noalign{\vspace {.05cm}}
\multirow{2}{*}{Crafted-9} & Actual & 27.2 & 12.1 & 5.9 & 3.6 &  47.5 & 106.6 & 217.3 & 358.0 \\
& Predicted & 27.0 & 13.5 & 6.7 & 3.3 & 48.0 & 96.0 & 192.0 & 384.0  \\
\hline
\hline
\end{tabular}
\caption{Parallel performance of Sparrow on crafted instances}
\label{table:sparrow-crafted-parallel}
\end{table}

Figure \ref{figure:craft-0-sparrow} shows the empirical and predicted performance of Sparrow to solve the instance Crafted-1. In particular, we would like to point out that the exponential distribution fits well the empirical data on 384 cores (Figure \ref{figure:crafted-0-sparrow-cdf-384cores}). On the other hand, Figure \ref{figure:crafted-0-sparrow-speedup}  shows, as expected, the predicted (ideal) linear speedup, and the speedup of the empirical data is also linear but with a slightly lower slope. In addition, the same behavior can be observed for the remaining instances, see Table \ref{table:sparrow-crafted-parallel} for complete results.

\subsection{Analysis}
Several works have been devoted to the experimental study of parallel multi-walk extensions of local search algorithms~\cite{arbelaez-ictai12,arbelaez-evocop13,Hoos-book}, but we presented in this paper the first approach (to our knowledge) which applies order statistics in order to predict the parallel performance of local search algorithms for SAT. Although most of the literature on runtime distributions uses the exponential distribution to estimate the theoretical performance of the parallel algorithm, results in Section \ref{section:experiments-random} show that it is sometimes more suitable to characterize the empirical runtime distribution by a lognormal or a shifted exponential distribution.

Interestingly, the phase transition point also seems to have important consequences in the parallel performance of local search algorithms. For Sparrow at least, which is the solver with an overall better speedup factor, the instances in the phase transition region are lognormally distributed, while instances outside the phase transition are shifted-exponentially distributed.
Another interesting aspect is that in theory the  probability  of returning  a  solution
in \emph{no} iterations is non-null because of the (uniform) random initialization.
However, in practice a minimum number of steps is in general required to reach a solution
cf. \cite{Hoos-rtd,tttplots2011} for the sequential case,
and therefore experimental data may be better approximated by a shifted distribution with $x_0>0$,
as it is the case in the random instances.
This leads to a non-linear speedup with a finite limit, even in the case of an exponential distribution.
Indeed, the experimental speedup for both CCASAT and Sparrow on random instances is far from linear.
On the contrary, Sparrow on crafted instances has a linear speedup which could be explained
by the fact that the  minimal runtime 
is negligible w.r.t. the mean time
(\ie $1/\lambda$ for an exponential distribution). Therefore, the statistical  test succeed
for $x_0 \simeq 0$. This suggests that, in general, the comparison between
the minimal time and the mean time is a key element for the study of the parallel behavior.

We do not discard that other parameters for the reference solvers would lead to other theoretical distributions (e.g. exponential distribution for random instances). In \cite{kroc-sat10} the authors showed that a well-tuned version of WalkSAT is exponentially distributed for instances in the phase transition region. However, we experimented by increasing the \textit{ps} (smoothing probability) parameter of Sparrow and still obtained the same theoretical distribution. In addition, when \textit{ps} is too high the solver was unable to solve the instances within the 3 hour time limit.  Unfortunately, CCASAT is only available in binary form, and it is not possible to experiment with other parameters for the solver.

We expect this work to have significant implications in the area of automatic parameter tuning to devise scalable local search algorithms. Currently, most parameter tuning tools (e.g. \cite{paramILS,paramGGA}) are designed to improve the expected mean (or median) runtime, however as observed in this paper, unless the algorithms exhibit a non-shifted exponential distribution, their parallel performance is far from linear  and varies from algorithm to algorithm.

\section{Conclusions and Future Work}
\label{section:conclusions}
This paper has presented a model to estimate and evaluate the performance of parallel local search algorithms for SAT.  This model, based on order statistics, predicts the parallel runtime execution of a given local search algorithm by analyzing the runtime distribution of its sequential version. Interestingly, we have observed that, for the two different algorithms and the variety of instances considered in this study,
the runtime distribution can be characterized using two types of distributions: exponential (shifted and non-shifted) and lognormal.

Extensive experimental results using the best local search solvers from the previous SAT competition, indicate that the model accurately matches the parallel performance of the empirical experiments up to 384 cores. Moreover, the theoretical model confirms
the empirical results reported in the literature for local search algorithms \cite{arbelaez-evocop13,Shylo11}
in showing that the best sequential local search solver is not always the best one in parallel settings.

A natural extension of this work would consist in estimating the parallel performance of a given algorithm for unseen instances, even without full sequential execution. To this end, we plan to combine the statistical model presented in this paper with the extensive literature for predicting the runtime a of a given sequential algorithm (see \cite{satzilla-journal}). In addition, we also plan to investigate the application of more (complex) distributions to characterize the distribution of other local search algorithms (e.g. CCASAT for crafted instances).


\bibliography{biblio}
\bibliographystyle{acmtrans}

\end{document}


\maketitle

\thispagestyle{myheadings}
\pagestyle{myheadings}

\begin{appendix}
\section{}

In this paper we consider a set of 369 instances from the SAT'11 competition (random category). Since our objective is to measure the performance improvement as the number of cores increases, we filtered out too easy and too hard instances by running Sparrow using 16 cores and used a subset of instances whose average runtime across 10 independent executions were between 2 and 3 minutes. This corresponds to about 30 minutes (average time) for the sequential algorithm.
The following list contains the final set of 10 instances (denoted Rand-[1 to 10]) in this paper):

\begin{itemize}
\item Rand-1: unif-k3-r4.2-v30000-c126000-S1854039067-041-UNKNOWN
\item  Rand-2: unif-k3-r4.2-v35000-c147000-S970100151-015-UNKNOWN
\item  Rand-3: unif-k3-r4.2-v40000-c168000-S1184456903-078-UNKNOWN
\item Rand-4: unif-k3-r4.2-v50000-c210000-S1170024351-015-UNKNOWN
\item Rand-5: unif-k3-r4.2-v50000-c210000-S537193780-078-UNKNOWN
\item Rand-6: unif-k3-r4.2-v50000-c210000-S957916968-041-UNKNOWN
\item Rand-7: unif-k5-r20-v2000-c40000-S922811046-076-UNKNOWN
\item Rand-8: unif-k5-r20-v1500-c30000-S976428817-077-UNKNOWN
\item Rand-9: unif-k5-r20-v1250-c25000-S573815729-001-UNKNOWN
\item Rand-10: unif-k5-r20-v2000-c40000-S1264065752-051-UNKNOWN
\end{itemize}

The second problem family consists of instances designed, either automatically or manually, to be difficult for SAT solvers. In particular we focus our attention on a set of 149 known SAT instances from the SAT'11 competition (crafted category). Again, we filtered out too easy and too hard instances by running in parallel 16 copies of Sparrow and selecting a subset of instances whose runtime across 10 independent executions were between 1 and 3 minutes. The following list includes the final set of 9 instances used in this paper (denoted crafted-[1 to 9] in this paper):

\begin{itemize}
\item crafted-1: srhd-sgi-m37-q505.75-n35-p15-s48276711
\item crafted-2: srhd-sgi-m42-q585-n40-p15-s54275047
\item crafted-3: srhd-sgi-m42-q663-n40-p15-s72490337
\item crafted-4: srhd-sgi-m47-q742.5-n45-p15-s28972035
\item crafted-5: em\_8\_4\_5\_fbc
\item crafted-6: rbsat-v1150c84314g7
\item crafted-7: rbsat-v1375c111739g4
\item crafted-8: sgen3-n240-s78945233-sat
\item crafted-9: sgen3-n260-s62321009-sat
\end{itemize}

\end{appendix}